\documentclass[aps,prl,onecolumn,superscriptaddress,10pt,floatfix,amsmath,amssymb]{revtex4-2}

\usepackage{graphicx} 
\usepackage{dcolumn}
\usepackage{xcolor}
\usepackage{bm}
\usepackage{physics}
\usepackage{lipsum}  
\usepackage{hyperref}
\usepackage[normalem]{ulem}
\graphicspath{{images/}}

\begin{document}

\title{On the application of a Silicon photomultiplier-based receiver for binary phase-shift-keying protocols}

\date{\today}

   \author{Silvia Cassina} 
   \affiliation{Department of Science and High Technology, University of Insubria, Via Valleggio 11, I-22100 Como (Italy)}

     \author{Michele N. Notarnicola}
   \affiliation{Department of Optics, Palack\'y University, 17 Listopadu 12, 771 46 Olomouc (Czech Republic) and University of Milan, Via Celoria 16, I-20133 Milan (Italy)} 
   
   \author{Stefano Olivares}
   \affiliation{University of Milan and INFN Section of Milan, Via Celoria 16, I-20133 Milan (Italy)} 
   
   \author{Alessia Allevi} \email{alessia.allevi@uninsubria.it}
   \affiliation{Como Lake Institute of Photonics, Department of Science and High Technology, University of Insubria, Via Valleggio 11, I-22100 Como (Italy)}
   \affiliation{Institute for Photonics and Nanotechnologies, IFN-CNR, Via Valleggio 11, I-22100 Como (Italy)}

\begin{abstract}
Over the past decade, binary phase-shift keying 
encoding has been used as a benchmark to test the performance of different detection strategies to address the problem of state discrimination. In this context, hybrid devices, giving access to both particle- and wave-like properties of light, could offer the possibility to better discriminate the sent signals. In this work, we consider a 
quantum channel exploiting a hybrid receiver embedding Silicon photomultipliers as photon-number-resolving detectors. We retrieve the discrimination error probability and the mutual information between sender and receiver as functions of some relevant experimental parameters in the case of binary phase-shifted coherent states. Our promising results, supported also with numerical simulations and theoretical analysis, foster further using this kind of hybrid receiver in more complex detections schemes.
\end{abstract}

\maketitle

\label{intro}
Binary-phase-shift-keying (BPSK) 
protocols represent one of the basic strategies to transmit information between two parties \cite{cariolaro,helstrom,bergou}, with the advantages of both a simple implementation scheme, realized by two coherent states phase-shifted by $\pi$, and more robustness to losses than nonclassical states \cite{izumi,muller}.
However, being different coherent states non orthogonal with one another, their perfect discrimination is not allowed by quantum mechanics laws \cite{becerra1,becerra2}. Therefore, over the years different strategies to reach the minimum error probability in the discrimination process, usually called Helstrom bound \cite{helstrom,bergou}, have been devised. Among them, it is worth to mention the optimal solution, represented by the Dolinar receiver, based on a direct detection scheme and adaptive displacements \cite{dolinar}. However, such a choice is quite difficult to realize from an experimental point of view as it requires fast closed-loop feedback control \cite{assalini}. A simpler feasible solution is provided by the Kennedy receiver, based on a nulling displacement operation followed by photodetection, which guarantees a less demanding implementation scheme, but with near optimum performance \cite{kennedy}. More generally, most of the strategies developed so far are based on the use of single-photon detectors that allow on-off detection \cite{flamini,sasaki}. 
However, it has been demonstrated that, in the presence of phase noise, direct detection schemes are no more optimal \cite{olivares13}.
To overcome this problem, DiMario and Becerra proposed the displacement photon-number-resolving (PNR) receiver \cite{dimario1}, that is a Kennedy setup employing PNR detectors with finite resolution instead of on-off photodetectors, with optimized decision rule and encoding strategy \cite{dimario2}.
An alternative is given by the use of a continuous variable detection system, namely homodyne detection \cite{lvovsky, grosshans}, that has been demonstrated to approach the Helstrom bound in the presence of large phase noise \cite{olivares13}.\\
Overall, we have essentially two well-established technologies to perform the detection: the PNR (or on-off, as limiting case) and the homodyne (or double-homodyne) ones \cite{olivaresPLA}. As a matter of fact, these are related to rather different detection apparatuses, including calibration, electronics and data post-processing. 
However, it is possible to merge the two technologies and obtain a hybrid receiver, called weak-field homodyne (WFH), that is particularly suitable for coherent states \cite{donati, allevi17}.
The detector is based on an interferometric scheme, like homodyne detection, but the typically-employed photodiodes are replaced by PNR detectors. Moreover, as a consequence of the detector sensitivity, the local oscillator (LO) is in the mesoscopic regime and not in the typical macroscopic one, meaning that its intensity is comparable to that of the signal \cite{njp}. This last difference between the two detection schemes is a consequence of the PNR detectors working range. We have already demonstrated that this scheme is quasi optimal in the case of BPSK coherent-states discrimination 
affected by uniform phase noise, similarly to what happens in the case of standard homodyne detection \cite{bina}.\\
Furthermore, some of us recently proved that WFH detection can be suitably combined with a displacement PNR scheme to obtain a more sophisticated receiver, the so-called HYNORE (hybrid near-optimum receiver), 
that beats the homodyne scheme and the Kennedy receiver, both in the presence and in the absence of phase noise, thus closing the gap with respect to the Helstrom bound \cite{notarnicola, notarnicolaff, notarnicolaphn}. Therefore, in view of realistic implementations, it is important to optimize the WFH scheme by properly choosing the class of PNR detectors to be used. To this aim, Silicon photomultipliers (SiPMs) provide a fascinating solution, since they are commercial detectors quite compact, endowed with a good PNR capability \cite{chesi,cassina} and, as we have recently shown in Ref.~\cite{OE24}, their application can be further extended to continuous-variable quantum key distribution (CVQKD), where the PNR capability 
is extremely useful to evaluate the secret key generation rate.\\
In this work, we focus on the problem of BPSK discrimination by exploring the conditions under which mutual information and error probability exhibit a complementary behavior. Using these two figures of merit, we investigate the role played by the intensity of the LO and the imbalance in the prior probability of sending the two coherent states. In this respect, we also consider the realistic case of an unbalanced beam splitter mixing signal and LO, which leads to a rapid decrease in both mutual information and error probability. 
The experimental results are in agreement with both numerical simulations and theoretical expectations, thus suggesting further exploitation of the WFH scheme in the HYNORE receiver.

\section{BPSK discrimination with SiPM: theoretical framework}
\label{theo}
In the framework of a coherent-state BPSK discrimination, the sender encodes two classical symbols $k=0,1$, generated with prior probabilities $q_k$, $q_0+q_1=1$, onto the coherent states:
\begin{equation}
    |\alpha_k\rangle = |\alpha  e^{i \phi_k} \rangle \, , \quad \alpha>0 \, ,
\end{equation}
with $\phi_0=\pi$ and $\phi_1=0$, such that $|\alpha_0\rangle = |-\alpha\rangle$ and $|\alpha_1\rangle=|\alpha\rangle$, respectively.
The pulses then reach the receiver, who
performs a positive-operator-valued measure (POVM) to infer the encoded symbol. In general, the receiver discriminates them with an error probability
\begin{equation}\label{eq:PerrGen}
P_{\rm err} = q_0\, p(1|0) + q_1\, p(0|1)\,,
\end{equation}
in which $p(i|j)$ is the conditional probability that the outcome $i$ is obtained when $j$ is sent.\\
In this paper, we consider a receiver based on a WFH scheme involving PNR detectors, such that the input state $|\alpha_k\rangle$, $k=0,1$, interferes at a balanced beam splitter (BS) with a low-intensity LO prepared in the coherent state $| z \rangle$, $z \in \mathbb{R}$. At each BS output PNR detection is performed and the photon-number difference $\Delta = n-m$, $\Delta \in \mathbb{Z}$, is evaluated in post processing. The distribution of the photon-number difference is given by the Skellam distribution \cite{skellam}: 
\begin{align}
S_{\Delta}(\mu_t,\mu_r)
= e^{-\mu_t - \mu_r} \left(\frac{\mu_t}{\mu_r}\right)^{\Delta/2}
I_{ \Delta}\big(2\sqrt{\mu_t \mu_r}\big) \, ,
\end{align}
where $I_{ \Delta}(x)$ is the modified Bessel function of the first kind and $\mu_{t(r)}$ is the mean photon number measured by the PNR detector at the transmitted (reflected) BS output, equal to:
\begin{subequations}\label{eq:mutmur}
\begin{align}
\mu_t(\phi_k) &=  \tau \alpha^2+ (1-\tau) z^2 + 2 \xi \, \sqrt{\tau(1-\tau)} \, z \alpha \, \cos{\phi_k} \,, \\[1ex]
\mu_r(\phi_k) &= (1-\tau) \alpha^2+ \tau  z^2 - 2 \xi \,\sqrt{\tau(1-\tau)} \, z \alpha \, \cos{\phi_k} \,, 
\end{align}
\end{subequations}
in which $\tau$ is the BS transmissivity, and $\xi \le 1$ is the interference visibility, that quantifies the spatial overlap between the signal and the LO beams interacting with each other \cite{banaszek}.
As we have already demonstrated in Ref.~\cite{bina}, this distribution converges to the homodyne distribution in the limit of high-intensity LO, i.e. $|z|^2 \gg 1$. 
\\
Assuming that $\tau=1/2$ and given the WFH outcome $\Delta \in \mathbb{Z}$, we construct the decision strategy in the more general case of unequal sampling, $q_0 \neq q_1$, according to the rule:
\begin{subequations}\label{eq:Decisionrule}
\begin{align}
\Delta <0 &\rightarrow \mbox{we infer state $|\alpha_0\rangle$}  \, , \\[1ex] 
\Delta >0 &\rightarrow \mbox{we infer state $|\alpha_1\rangle$}  \, ,  \\[1ex]
\Delta =0 &\rightarrow \mbox{random decision: we infer $|\alpha_k\rangle$ with probability $q_k$} \, ,
\end{align}
\end{subequations}
so that the corresponding error probability of the WFH receiver reads:
\begin{align}\label{perrtheo}
P_{\rm err} &= q_0\, \left[\sum_{\Delta=1}^{\infty} S_{\Delta}\big(\mu_t(0),\mu_r(0)\big)+ q_1 S_0 \right]  \nonumber \\
&\hspace{2.cm}+ q_1\, \left[\sum_{\Delta=-\infty}^{-1} S_{\Delta}\big(\mu_t(\pi),\mu_r(\pi)\big) + q_0 S_0 \right] \, ,
\end{align}
where $S_0= S_{\Delta=0}(\mu_t(0),\mu_r(0))= S_{\Delta=0}(\mu_t(\pi),\mu_r(\pi))$ is the value of the Skellam distribution in the case of inconclusive measurements.\\

As a counterpart to the error probability, we also evaluate the mutual information (MI) associated with the receiver, that provides the typical figure of merit to quantify the information transmission in quantum communication protocols, embedded with suitable codewords and reconciliation procedures. However, we underline that the strategy in Eq.~(\ref{eq:Decisionrule}) does not represent the optimal decoding strategy in a communication scenario, since it ultimately results in a binary decision rule \cite{OE24}. Therefore, the purpose of this analysis is simply to test the complementary behaviour of error probability and MI in a discrimination protocol, without addressing the role of WFH for communication \cite{WFHComm}.
\\
Generally speaking, for any communication channel described by two random correlated variables $X$ and $Y$, with prior and conditional probability distributions $p_X(x)$ and $p_{Y|X}(y|x)$, respectively, the MI is defined as: \cite{cover}
\begin{align}\label{eq:MI}
{\cal I} &= H(Y) - H(Y|X) \, ,
\end{align}
where $H(Y)=-\sum_y p_Y(y) \log_2 p_Y(y)$ and $H(Y|X)=-\sum_x p_{X}(x) \cdot\\
\sum_y p_{Y|X}(y|x) \times \log_2 p_{Y|X}(y|x)$ are the overall Shannon entropy of $Y$ and the average Shannon conditional entropy of $Y$ given $X$, respectively. The expression~(\ref{eq:MI}) quantifies the amount of correlations between the input and output channel variables and, according to Shannon's coding theorem, it represents the maximum information rate extractable from the channel \cite{cover}.
\\
Given this premise, in the case under investigation, namely WFH binary receiver with BPSK encoding, we deal with a binary input variable $k=0,1$ generated with prior probability distribution $\{q_0,q_1\}$, and a binary output $j=0,1$ derived by the decision rule~(\ref{eq:Decisionrule}). The conditional probability $p(j|k)$ then reads:
\begin{align}\label{eq:CondP}
p(0|k)=\sum_{\Delta=-\infty}^{-1} S_{\Delta}(\mu_t(\phi_k),\mu_r(\phi_k)) + q_0 S_0 \, , \nonumber\\[1ex]
p(1|k)=\sum_{\Delta=1}^{\infty} S_{\Delta}(\mu_t(\phi_k),\mu_r(\phi_k)) + q_1 S_0 \, ,
\end{align}
from which we construct the overall output distribution as $p(j)=\sum_k q_k p(j|k)$. Finally, the MI is obtained as:
\begin{align}\label{MItheo}
{\cal I} &= \sum_{j,k=0,1} q_k \, p(j|k) \log_2 \left( \frac{p(j|k)}{p(j)} \right) \nonumber \\[1ex] 
&= q_0\left[ p(0|0) \log_2 \left( \frac{p(0|0)}{p(0)} \right) + p(1|0) \log_2 \left( \frac{p(1|0)}{p(1)} \right)\right] \nonumber\\[1ex]
& \hspace{1cm} + q_1 \left[ p(0|1) \log_2 \left( \frac{p(0|1)}{p(0)}\right) + p(1|1) \log_2 \left( \frac{p(1|1)}{p(1)} \right) \right] \, .
\end{align}
 
The behavior of error probability and MI can be evaluated as functions of different parameters. In view of practical BPSK communication, it is worth experimentally investigating the possible limitations imposed by the use of a mesoscopic LO, that is $|z|^2\lesssim 15$, and the effect of an imbalance in the prior probability under conditions of unbalanced BS transmissivity.

\section{BPSK discrimination with SiPM: the experiment}
\label{exp}
\begin{figure}[t]
\centering
\includegraphics[width=13cm]{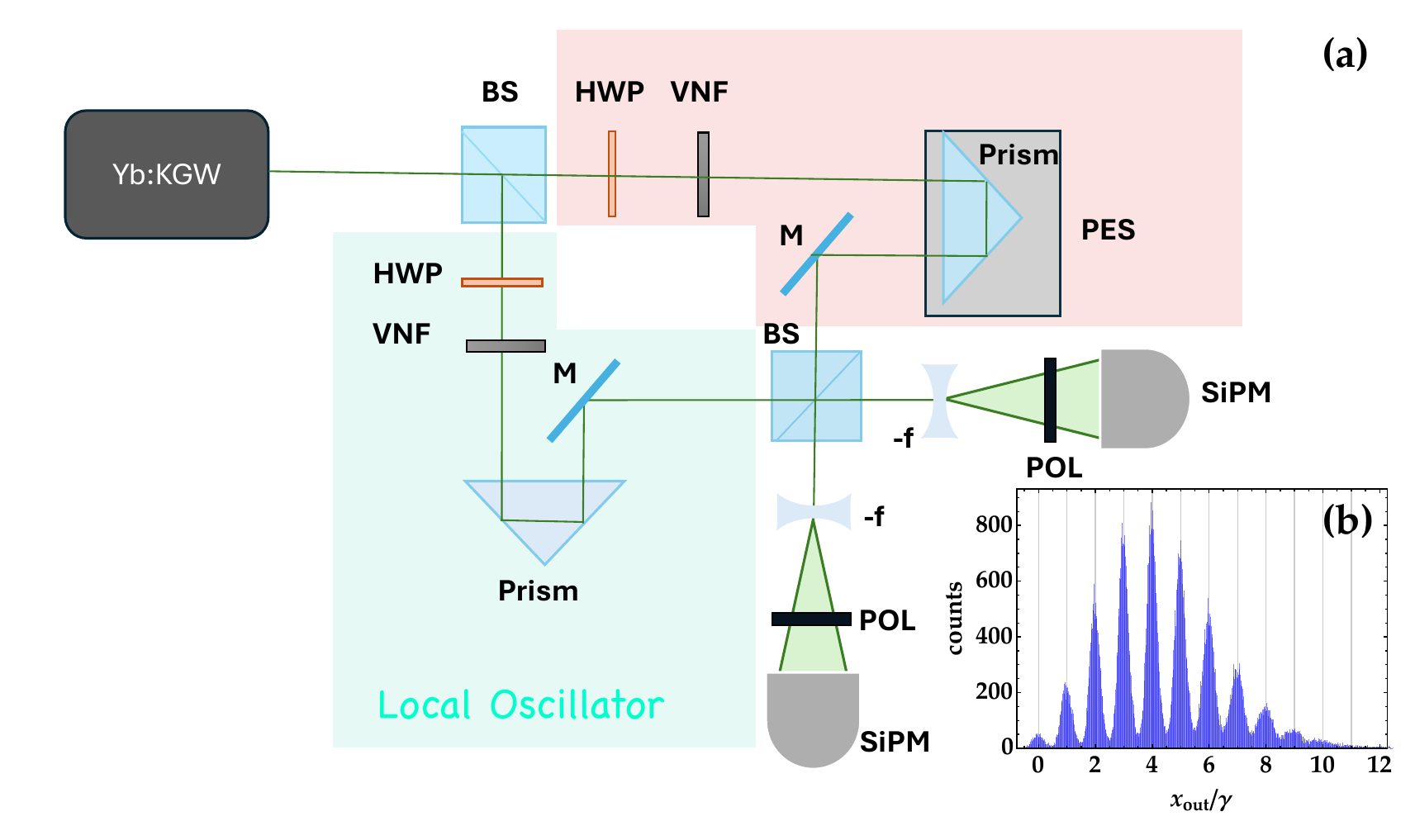}
\caption{(a) Sketch of the experimental setup, in which BS is a beam splitter, HWP is the half-wave plate, VNF is a variable neutral filter, M is a mirror, PES is the piezoelectric stage, -f is a divergent lens, POL is a polarizer, and SiPM is a Silicon photomultiplier. (b) Typical pulse height spectrum in the case of $\langle m \rangle = 4.5$.}\label{setup}
\end{figure}

\begin{figure}[t]
\centering
\includegraphics[width=14cm]{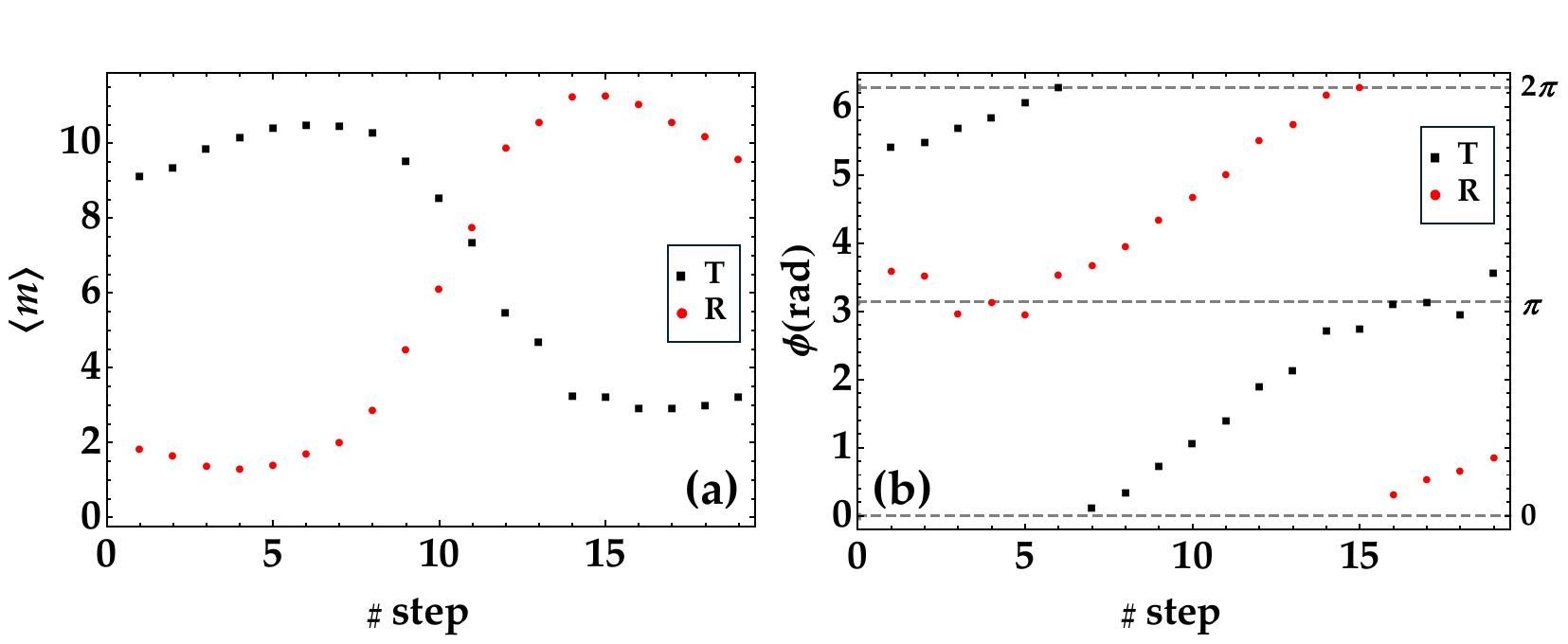}
\caption{(a) Mean number of photons measured on the transmitted (T, black dots) and reflected (R, red dots) arms of the interferometer as a function of the number of steps of the piezoelectric movement. (b) Retrieved phase as a function of the number of steps.}\label{step}
\end{figure}

\begin{figure}[t]
\centering
\includegraphics[width=14cm]{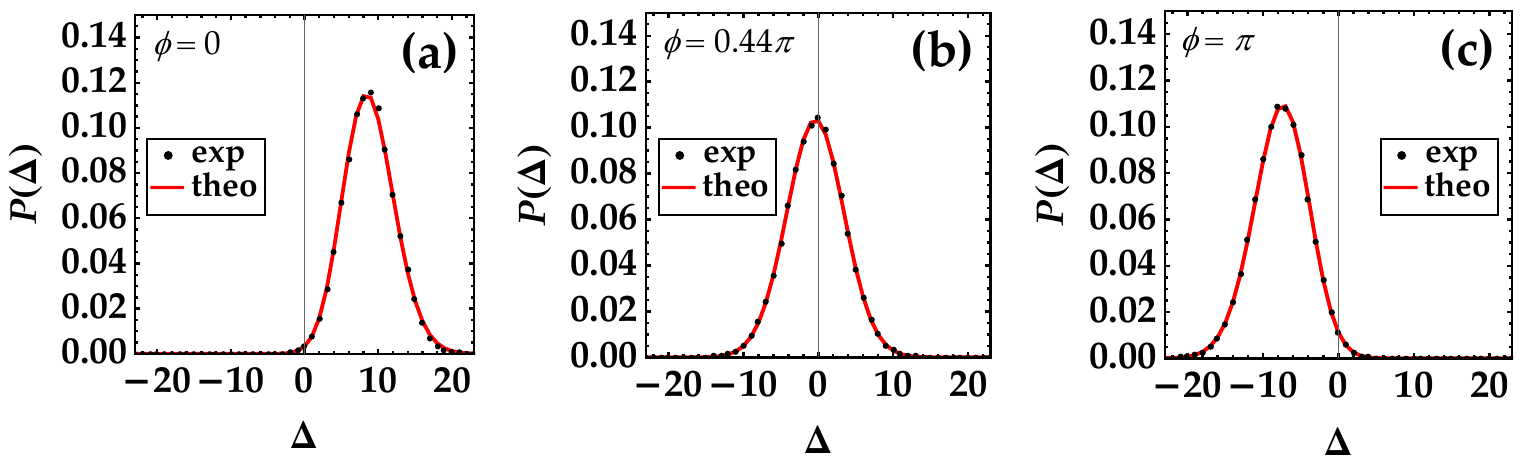}
\caption{(a)-(c): Distribution of the photon-number difference for detected photons in the case of $\phi=0$ rad, $\phi=0.44 \pi$ rad, and $\phi=2 \pi$ rad, respectively. Black dots: experimental data, red line: reconstructed Skellam distribution for $|\alpha|^2 = 2.37$ and $|z|^2 = 10.88$.}\label{skellam}
\end{figure}

The realized setup is shown in Fig.~\ref{setup}(a): the second-harmonic pulses (at 515 nm, pulse duration of 190 fs) of a Yb:KGW laser operated at 5 kHz are sent to a Mach-Zehnder interferometer. Its first beam splitter divides the light into two parts: the transmitted beam represents the signal, while the reflected one the LO. On each arm a variable neutral filter (VNF) is used to vary the intensity of light. To avoid different attenuations of the different portions of the beam spot, before entering the interferometer, the size of the beam has been reduced through a telescope. Moreover, the polarization has been kept under control by means of two half-wave plates placed just behind the first beam splitter and two polarizers at the exit of the interferometer. Its two outputs are spatially selected by means of two apertures, collected by two multi-mode fibers (1-mm core diameter) and delivered to two PNR detectors. In more detail, the hybrid receiver we built is based on the use of two SiPMs, which are commercial detectors characterized by a good photon-number-resolving capability. This is well visible in the pulse-height spectrum shown in panel (b) of Fig.~\ref{setup}. 
The model we use is the MPPC S13360-1350CS produced by Hamamatsu Photonics \cite{hama}, which consists of 667 pixels in a 1.3 $\times$ 1.3 mm$^2$ photosensitive area, with a pixel pitch equal to 50 $\mu$m. Each detector output is amplified by a fast amplifier embedded in the computer-based Caen SP5600 Power Supply and Amplification Unit, synchronously integrated by a boxcar-gated integrator, and digitized \cite{scirep19}. The integration of the signal corresponding to each detector over a small gate width (15 ns) allows us to make the dark count rate almost negligible (0.3$\%$) \cite{chesi,cassina}.\\ 
To produce the states $| \pm \alpha \rangle$, the translation stage on which the piezo is mounted is moved in steps for each of which $10^5$ consecutive pulses are acquired. 
By applying the self-consistent method \cite{JMO, APL} to the outputs of the detection chain, it is possible to reconstruct the statistical properties of light. For instance, it is possible to calculate, for each detector, the mean value of light as a function of the piezolelectric stage (PES). A typical behavior is shown in Fig.~\ref{step}(a) for the detector placed in the transmitted arm (black dots) and that in the reflected arm (red dots). 
As already explained in Ref.~\cite{JOSAB10, scirep16}, from the mean value it is also possible to extract information about the relative phase between signal and LO, as shown in Fig.~\ref{step}(b). The same kind of information could be obtained by plotting the photon-number difference between the photons measured in the two arms as a function of the step of the piezoelectric stage.\\ 
Typical distributions of the photon-number difference are shown in Fig.~\ref{skellam}, 
where three different conditions are presented: in panel (a) there is the case corresponding to $|- \alpha \rangle$, in panel (b) that roughly corresponding to $| e^{i \pi/2} \alpha \rangle$, and in panel (c) that corresponding to $| \alpha \rangle$. For the sake of completeness, in each plot the data are superimposed on the expected Skellam distribution.\\
These strategies allow us to properly select the states $|\alpha \rangle$ and $|- \alpha \rangle$ for the different conditions we investigated.
The experimental results are also compared to those obtained by numerical simulation. By assuming a given prior probability, we randomly generate either $|\alpha \rangle$ or $|- \alpha \rangle$ with a fixed mean value and simulate its superposition, with visibility $\xi$, on a LO with fixed mean value at a BS with transmittance $\tau$. We calculate the shot-by-shot photon number difference between the two BS outputs and use it to calculate both the error probability and MI. For each choice of the mentioned parameters, we consider three sets of $5 \cdot 10^4$ data, from which the error bars are also obtained. The introduction of the overlap value aims to simulate the contribution of this effect to the explored quantities, as well as to better compare the simulated data to those obtained from the experiment. In fact, in a free space setup the contribution of an imperfect overlap is non-negligible \cite{OE24}, as shown hereafter in the data analysis. A fiber-based setup could make this problem negligible, but at this stage, it is interesting to explore the role it plays on the two figures of merit.

\subsection{Assessing the impact of a low-intensity LO}

\begin{figure}[t]
\centering
\includegraphics[width=14cm]{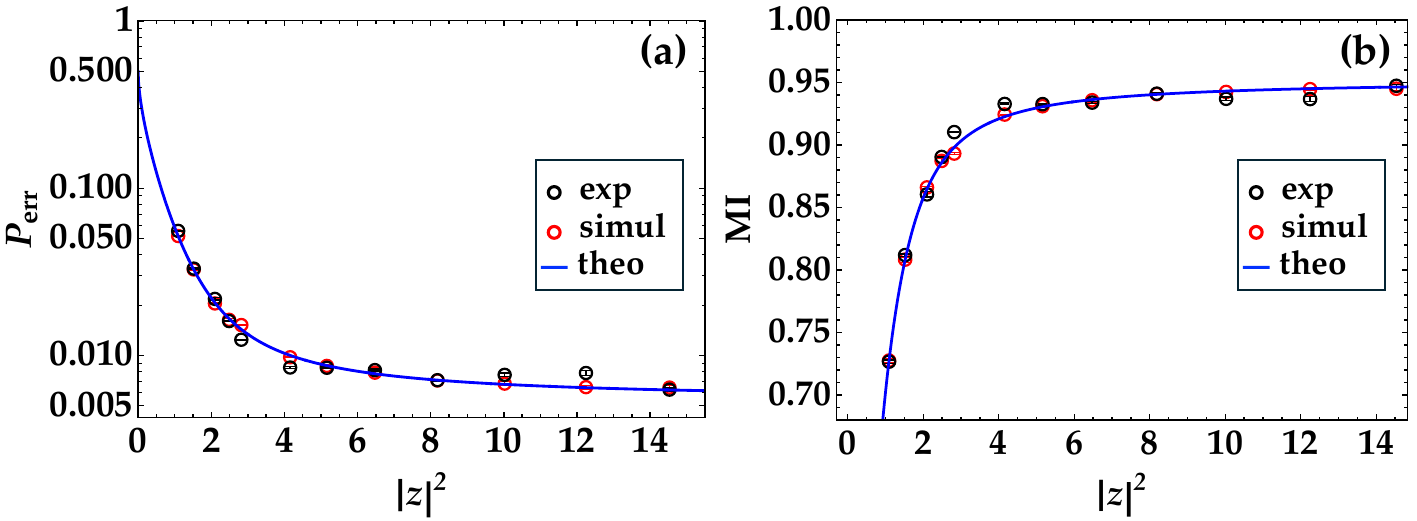}
\caption{(a) and (b): Error probability and mutual information as a function of $|z|^2$ for $|\alpha|^2 = 1.97$ in the case of $q_0 = q_1 = 1/2$. Black circles: experimental data, red circles: simulated data, in which $\xi = 0.91$ and $\tau = 0.5$; blue curve: theoretical expectations according to Eqs.~(\ref{perrtheo}) and (\ref{MItheo}) with the same values of the parameters used in the simulations.}\label{FMvsLO}
\end{figure}

As a first study, we monitor the possible limitations introduced by the mesoscopic LO evaluating the error probability and MI as functions of the mean value of LO. 
In Fig.~\ref{FMvsLO}(a) we show the results obtained for the error probability, while in panel (b) of the same figure we plot the ones related to MI. As it can be noticed, the two panels present a complementary behavior: increasing the LO the error probability decreases, while the mutual information increases. Moreover, both quantities exhibit an asymptotic behavior, proving that an infinite LO is not necessary. This represents a strong result compared to the requirement of a macroscopic LO for the case of standard homodyne receivers.\\  
In both panels, black circles correspond to experimental data, while red circles correspond to simulated data, in which we used the same values of signal and LO as those experimentally measured and assumed an overlap equal to $\xi = 0.91$ and a transmittance $\tau = 0.5$. The theoretical expectations according to Eqs.~(\ref{perrtheo}) and (\ref{MItheo}) with the same values of the parameters used in the simulations are superimposed on the data.\\

\subsection{Robustness with respect to unequal sampling and unbalanced BS}

As a second study, we test the robustness of the receiver for state discrimination applications by considering both the cases of unequal prior probabilities $q_0\ne q_1$, and imperfect detection due to unbalanced transmissivity of the BS in the WFH scheme of Fig.~\ref{setup}. In this scenario, we fix the mean value of LO, $|z|^2$, and a BS transmissivity $\tau \ne 0.5$, and study the behavior of both the error probability and MI as a function of the prior probability $q_0>1/2$ of generating state $|\alpha_0\rangle$.
\begin{figure}[t]
\centering
\includegraphics[width=14cm]{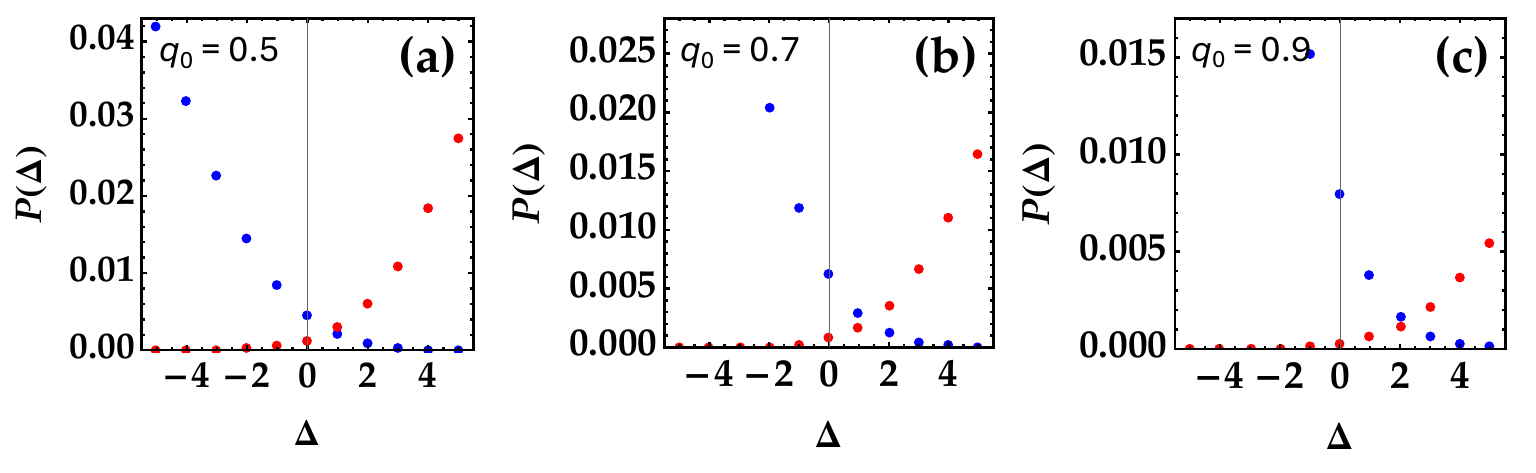}
\caption{(a)-(c): $q_0 S_{\Delta}\big(\mu_t(0),\mu_r(0)\big)$, corresponding to the state $|-\alpha \rangle$ (blue dots), and $q_1 S_{\Delta}\big(\mu_t(\pi),\mu_r(\pi)\big)$, corresponding to the state $|\alpha \rangle$ (red dots), in the case of $q_0 = 0.5$, $q_0 = 0.7$, and $q_0 = 0.9$, respectively. The other parameters are: $|\alpha|^2 = 2.37$, $|z|^2 = 10.88$, $\xi = 0.86$, and $\tau = 0.545$.}\label{skellam_q0}
\end{figure}
Remarkably, as depicted in Fig.~\ref{skellam_q0}, when $\tau \ne 0.5$, the outcome $\Delta=0$ is no longer inconclusive, as $S_{\Delta=0}(\mu_t(0),\mu_r(0))\ne S_{\Delta=0}(\mu_t(\pi),\mu_r(\pi))$, see Eqs.~(\ref{eq:mutmur}). In turn, the decision rule~(\ref{eq:Decisionrule}) should be properly modified according to the maximum a posteriori probability (MAP) criterion: that is, for a given $\Delta \in \mathbb{Z}$, we infer the state $|\alpha_k\rangle$ with the largest a posteriori probability according to Bayes theorem, namely:
\begin{align}
    {\cal P}_{\rm post}(k|\Delta)= \frac{q_k S_\Delta(\mu_t(\phi_k),\mu_r(\phi_k))}{\cal N} \, ,
\end{align}
${\cal N}$ being a suitable normalization factor
\cite{notarnicola, notarnicolaff,notarnicolaphn}. That is, we perform the decision ``0" whenever $q_0 S_\Delta(\mu_t(\phi_0),\mu_r(\phi_0))> q_1 S_\Delta(\mu_t(\phi_1),\mu_r(\phi_1))$, and vice versa.
This is equivalent to introducing a threshold outcome dependent on prior probability $\Delta_{\rm th}=\Delta_{\rm th}(q_0)$ such that if $\Delta \le \Delta_{\rm th}$ we infer the state $|\alpha_0\rangle$, while if $\Delta > \Delta_{\rm th}$ we infer $|\alpha_1\rangle$. Accordingly, the conditional probability $p(j|k)$ in~(\ref{eq:CondP}) is changed into:
\begin{align}
p_{\rm unb}(0|k)&=\sum_{\Delta=-\infty}^{\Delta_{\rm th}} S_{\Delta}(\mu_t(\phi_k),\mu_r(\phi_k)) \, , \nonumber\\[1ex]
p_{\rm unb}(1|k)&=\sum_{\Delta=\Delta_{\rm th}+1}^{\infty} S_{\Delta}(\mu_t(\phi_k),\mu_r(\phi_k))\, ,
\end{align}
from which we compute the error probability $P_{\rm err}$ and the MI thanks to Eqs.~(\ref{eq:PerrGen}) and~(\ref{MItheo}), respectively.
\begin{figure}[t]
\centering
\includegraphics[width=14cm]{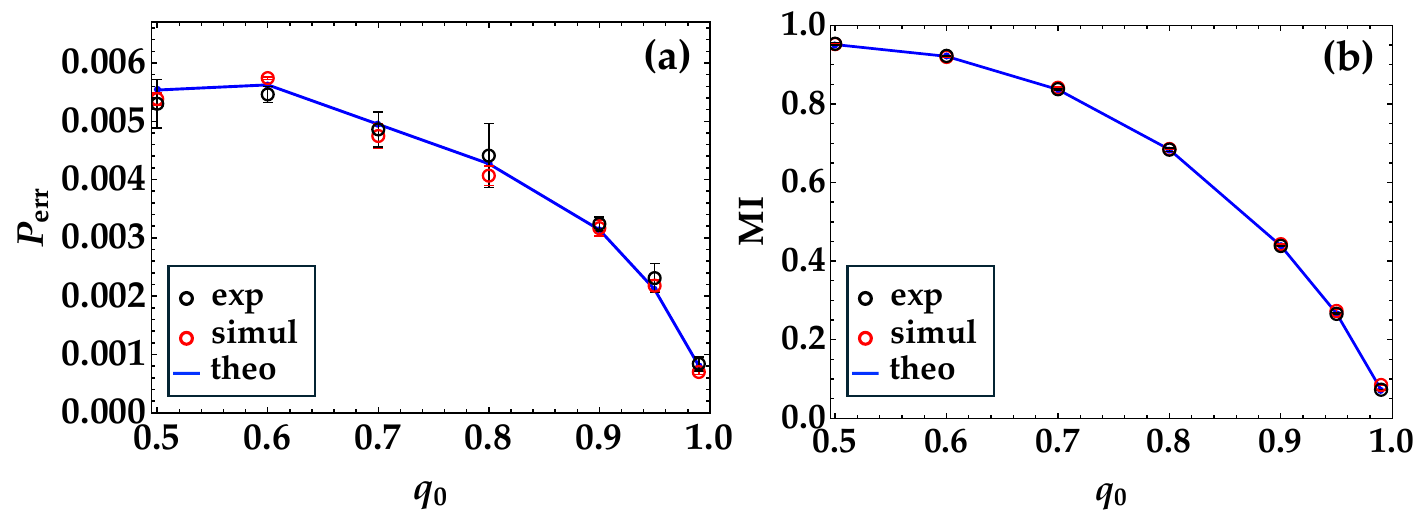}
\caption{(a) and (b): Error probability and mutual information as a function of the prior probability, $q_0$. Black circles: experimental data, red circles: simulated data for $|\alpha|^2 = 2.37$, $|z|^2 = 10.88$, $\xi = 0.86$, and $\tau = 0.545$; blue curve: theoretical expectations according to Eqs.~(\ref{perrtheo}) and (\ref{MItheo}) with the same values of the parameters used in the simulations.}\label{FMvsp}
\end{figure}
The data are presented in Fig.~\ref{FMvsp}(a) and (b) for error probability and MI, respectively.
As it can be noticed, in this case the two quantities exhibit an almost similar behavior. In fact, varying the prior probability is detrimental in the case of MI, since the amount of information shared by the two parties decreases as the sent state becomes more certain. At the same time, increasing the value of $q_0$ decreases the error probability since it is easier to discriminate which state has been sent if one of the two states is sent with less probability than the other. Our statement is corroborated by the numerical simulations and the theoretical expectation evaluated in the experimental values of the parameters. It is worth noting that the non-trivial behavior obtained in this case proves that the theoretical model used to describe the employed receiver is correct. Moreover, our hybrid scheme seems to be useful to deal with the problem of state discrimination, thus suggesting that further investigations in more complex setups, such as the HYNORE, could be desirable. 

\section{Conclusions}
In this work, we investigated the performance of a SiPM-based hybrid receiver to address BPSK communication with coherent states having a $\pi$- difference in phase. This preliminary analysis represents a fundamental steps towards the implementation of a more complex detection system, such as the HYNORE receiver, in which our scheme can be embedded. To fully characterize the receiver, we explored both the behavior of error probability and mutual information as functions of some relevant parameters. First of all, we investigated the two figures of merit as functions of the energy of the LO, proving that in our receiver a macroscopic LO is not necessary. Secondly, we studied the behavior of MI and error probability as functions of the prior probability. Interesting results are also achieved in this case, showing the criticality of the parameters involved in the description of the physical system, such as the transmissivity of the BS at which signal and LO are mixed.
This analysis suggests the further exploitation of the system in a more complex network as well as for different application fields.

\section{Funding}
S.C. and A.A. acknowledge the support by PNRR D.D.M.M. 351/2022. 

\section{Disclosures}
The authors declare no conflicts of interest.

\section{Data Availability Statement}
The datasets generated and analyzed during the current study are available from the corresponding authors on reasonable request.


\section{References}
\begin{thebibliography}{00}
\bibitem{cariolaro} G. Cariolaro, Quantum Communications, Springer Publishing Company Incorporated, Switzerland (2015).
\bibitem{helstrom} C. W. Helstrom, Quantum Detection and Estimation Theory, Elsevier, Academic Press (1976).
\bibitem{bergou} J. A. Bergou, Discrimination of quantum states, J. Mod. Opt. 57 (2010) 160-180.
\bibitem{izumi} S. Izumi, M. Takeoka, M. Fujiwara, N. Dalla Pozza, A. Assalini, K. Ema, M. Sasaki, Displacement receiver for phase-shift-keyed coherent states, Phys. Rev. A 86 (2012) 042328.
\bibitem{muller} C. R. M$\ddot{\rm u}$ller, Ch. Marquardt, A robust quantum receiver for phase shift keyed signals, New J. Phys. 17 (2015) 032003.
\bibitem{becerra1} F. E. Becerra, J. Fan, G. Baumgartner, J. Goldhar, J. T. Kosloski, A. Migdall, Experimental demonstration of a receiver beating the standard quantum limit for multiple nonorthogonal state discrimination, Nat. Photon. 7 (2013) 147–152.
\bibitem{becerra2} F. E. Becerra, J. Fan, A. Migdall, Implementation of generalized quantum measurements for unambiguous discrimination of multiple non-orthogonal coherent states, Nat. Commun. 4 (2013) 2028.
\bibitem{dolinar} S. J. Dolinar, An optimum receiver for the binary coherent state quantum channel, Massachusetts Institute of Technology, Cambridge, Technical Report (1973).
\bibitem{assalini} A. Assalini, N. Dalla Pozza, G. Pierobon, Revisiting the Dolinar receiver through multiple-copy state discrimination theory, Phys. Rev. A 84 (2011), 022342.
\bibitem{kennedy} R. S. Kennedy, A Near-Optimum Receiver for the Binary Coherent State Quantum Channel, Quarterly Progress Report 108 (1973) 219-225.
\bibitem{flamini} F. Flamini, N. Spagnolo, F. Sciarrino, Photonic quantum information processing: a review, Rep. Prog. Phys. 82 (2019) 016001.
\bibitem{sasaki} M. Sasaki, O. Hirota, Optimum decision scheme with a unitary control process for binary quantum-state signals, Phys. Rev. A 54 (1996) 2728-2736.
\bibitem{olivares13} S. Olivares, S. Cialdi, F. Castelli, M. G. A. Paris, Homodyne detection as a near-optimum receiver for phase-shift-keyed binary communication in the presence of phase diffusion, Phys. Rev. A 87 (2013) 050303(R).
\bibitem{dimario1} M. T. DiMario, F. E. Becerra, Robust Measurement for the Discrimination of Binary Coherent States, Phys. Rev. Lett. 121 (2018) 023603.
\bibitem{dimario2} M. T. DiMario, L. Kunz, K. Banaszek, F. E. Becerra, Optimized communication strategies with binary coherent states over phase noise channels, npj Quantum Inf. 5 (2019) 65.
\bibitem{lvovsky} A. I. Lvovsky, M. G. Raymer, Continuous-variable optical quantum-state tomography, Rev. Mod. Phys. 81 (2009) 299-332.
\bibitem{grosshans} F. Grosshans, P. Grangier, Continuous Variable Quantum Cryptography Using Coherent States, Phys. Rev. Lett. 88 (2002) 057902.
\bibitem{olivaresPLA} S. Olivares, Introduction to generation, manipulation and characterization of optical quantum states, Phys. Lett. A 418 (2021) 127720
\bibitem{donati} G. Donati, T. J. Bartley, X.-M. Jin, M.-D. Vidrighin, A. Datta, M. Barbieri, I. A. Walmsley, Observing optical coherence across Fock layers with weak-field homodyne detectors, Nature Commun. 5 (2014) 5584.
\bibitem{allevi17} A. Allevi, M. Bina, S. Olivares, M. Bondani, Homodyne-like detection scheme based on photon-number-resolving detectors, Int. J. Quantum Inform. 15 (2017) 1740016.
\bibitem{njp} S. Olivares, A. Allevi, G. Caiazzo, M. G. A. Paris, M. Bondani, Quantum tomography of light states by photon-number-resolving detectors, New J. Phys. 21 (2019) 103045.
\bibitem{bina} M. Bina, A. Allevi, M. Bondani, S. Olivares, Homodyne-like detection for coherent state-discrimination in the presence of phase noise, Opt. Express, 25 (2017) 10685-10692.
\bibitem{notarnicola} M. N. Notarnicola, M. G. A. Paris, S. Olivares, Hybrid near-optimum binary receiver with realistic photon- number-resolving detectors, J. Opt. Soc. Am. B 40 (2023) 705-714.
\bibitem{notarnicolaff} M. N. Notarnicola, S. Olivares, Beating the standard quantum limit for binary phase-shift-keying discrimination
with a realistic hybrid feed-forward receiver, Phys. Rev. A 108 (2023) 042619.
\bibitem{notarnicolaphn} M. N. Notarnicola, S. Olivares, A robust hybrid receiver for binary phase-shift keying discrimination in the presence of phase noise, Int. J. Quantum Inf. 22 (2024) 2450008.
\bibitem{chesi} G. Chesi, L. Malinverno, A. Allevi, R. Santoro, M. Caccia, M. Bondani, Measuring nonclassicality with silicon photomultipliers, Opt. Lett. 44 (2019) 1371-1374.
\bibitem{cassina} S. Cassina, A. Allevi, V. Mascagna, M. Prest, E. Vallazza, M. Bondani, Exploiting the wide dynamic range of silicon photomultipliers for quantum optics applications, EPJ Quantum Technol. 8 (2021) 4.
\bibitem{OE24} A. Sanvito, S. Cassina, M. Lamperti, M. N. Notarnicola, S. Olivares, A. Allevi, Assessing a binary quantum channel through a SiPM-based weak-field homodyne receiver, Opt. Express 32 (2024) 39846-39859.
\bibitem{skellam} J. G. Skellam, The frequency distribution of the difference between two Poisson variates belonging to different
populations, J. Roy. Statist. Soc. (N. S.) 109 (1946) 296.
\bibitem{banaszek} K. Banaszek, A. Dragan, K. W${\rm \acute{o}}$dkiewicz, and C. Radzewicz, Direct measurement of optical quasi-distribution functions: Multimode theory and homodyne tests of Bell’s inequalities, Phys. Rev. A 66 (2002) 043803.
\bibitem{WFHComm} M. N. Notarnicola, S. Olivares, Employing weak-field homodyne detection for quantum communications, arXiv:2405.14310 [quant-ph] (2024).
\bibitem{cover} T. M. Cover, Elements of information theory (John Wiley Sons, 1999).
\bibitem{hama} https://www.hamamatsu.com/content/dam/hamamatsu-photonics/sites/documents/99 SALES LIBRARY/ssd/s13360 series kapd1052e.pdf.
\bibitem{scirep19} G. Chesi, L. Malinverno, A. Allevi, R. Santoro, M. Caccia, A.  Martemiyanov, M. Bondani, Optimizing Silicon photomultipliers for Quantum Optics, Sci. Rep. 9 (2019) 7433.
\bibitem{JMO} M. Bondani, A. Allevi, A. Agliati, A. Andreoni, Self-consistent characterization of light statistics, J. Mod. Opt. 56 (2009) 226-231.
\bibitem{APL} J. Pe$\check{\rm r}$ina Jr., O. Haderka, A. Allevi, M. Bondani, Absolute calibration of photon-number-resolving detectors with
an analog output using twin beams, Appl. Phys. Lett. 104 (2014) 041113.
\bibitem{JOSAB10} M. Bondani, A. Allevi, A. Andreoni, Self-consistent phase determination for Wigner function reconstruction, J. Opt. Soc. Am. B 27 (2010) 333-337.
\bibitem{scirep16} M. Bina, A. Allevi, M. Bondani, S. Olivares, Phase-reference monitoring in coherent-state discrimination assisted by a photon-number resolving detector, Sci. Rep. 6 (2016) 26025.

\end{thebibliography}
\end{document}